\documentclass[11pt]{article}
\usepackage[a4paper,margin=2.3cm]{geometry}
\usepackage[utf8]{inputenc}
\usepackage{amsmath,amssymb,amsfonts}
\usepackage{bm}
\usepackage{graphicx}
\usepackage{xcolor}
\usepackage{authblk}
\usepackage{lineno}         
\usepackage{caption}
\usepackage[hidelinks]{hyperref}
\usepackage[super,comma,sort&compress]{natbib}
\usepackage{booktabs}
\newcommand{\NbSe}{NbSe$_2$}
\newcommand{\misfit}{(La$_x$Pb$_{1-x}$Se)$_{1.14}$(NbSe$_2$)$_2$}
\newcommand{\Tc}{T_{\mathrm{c}}}
\newcommand{\dvec}{\mathbf{d}}

\newcommand{\nM}{n_{M}}

\title{\bfseries Doping-controlled topological superconducting transition in misfit layer compounds}

\author[1,$\ast$]{Hugo Le Du}
\author[1]{Robin Salvatore}
\author[2]{Justine Cordiez}
\author[3]{Ludovica Zullo}
\author[1]{Arindam Mukherjee}
\author[1]{Daniel Schmieg}
\author[4]{Dominik Volavka}
\author[1]{François Debontridder}
\author[1]{Marie Hervé}
\author[4]{Tomas Samuely}
\author[2]{Shunsuke Sasaki}
\author[2]{Florent Pawula}
\author[2]{Etienne Janod}
\author[2]{Laurent Cario}
\author[1,$\ast$]{Tristan Cren}

\affil[1]{\small Institut des Nanosciences de Paris, Paris, France}
\affil[2]{\small Institut des Matériaux de Nantes Jean Rouxel, Nantes, France}
\affil[3]{\small Institut f{\"u}r Theoretische Physik und Astrophysik and W{\"u}rzburg-Dresden Cluster of Excellence ctd.qmat, Universität W{\"u}rzburg, 97074 W{\"u}rzburg, Germany}
\affil[4]{\small Centre of Low Temperature Physics, Pavol Jozef Šafárik University, Košice, Slovakia}
\affil[$\ast$]{\small Corresponding authors: hugo.ledu@sorbonne-universite.fr, tristan.cren@sorbonne-universite.fr}

\date{}

\begin{document}
\maketitle

\begin{abstract}
\noindent Achieving topological superconductivity is a key goal in quantum physics, offering a path to fault-tolerant quantum computers. A central challenge in this field is to continuously drive a material through a topological quantum phase transition to directly observe the evolution from  trivial to topological superconductivity. However, finding a robust platform that allows such extreme and precise tuning remains a challenge. Here, we demonstrate a doping-controlled phase transition from a conventional to a topological superconducting state in the bulk misfit layer compound \misfit. We reveal a non-monotonic phase diagram characterized by two distinct superconducting regimes separated by a non-superconducting phase at a precise doping. In the highly doped regime, the superconducting phase becomes remarkably sensitive to non-magnetic disorder, and orientation-selective in-gap modes emerge at atomic step edges. Supported by Bogoliubov-de Gennes calculations, these emergent spatial signatures are consistent with a time-reversal-symmetric crystalline-topological order parameter. Our results establish misfit compounds as a platform to engineering topological superconductivity.

\end{abstract}

\vspace{0.5em}

\section*{Introduction}

Driving a solid-state system through a topological quantum phase transition is a central goal for realizing fault-tolerant quantum computation \cite{Kitaev_2001, Nayak_2008}. Experimental signatures of topological superconductivity have been reported in a handful of platforms, from single atomic layers of Pb coupled to magnetic islands~\cite{Menard_2017} to the topological surface states of iron-based superconductors~\cite{Wang_2018}. A common ingredient is an external control parameter, a knob, that drives the material across the transition. The most popular road relies on applying a strong external magnetic field to one-dimensional proximitized semiconductor nanowires~\cite{Lutchyn_2010, Oreg_2010, Mourik_2012}. Compositional alloying of iron-based superconductors~\cite{Zhang_observation_2018} was also used to induce superconductivity in topological surface states, while the bulk remains a trivial superconductor. More recently, rhombohedral tetralayer and pentalayer graphene have been shown to host, in gate-induced flat bands, a superconducting state with signatures of chiral pairing~\cite{Han_2025}. Achieving a continuously tunable transition in a purely two-dimensional platform remains a formidable challenge. In 2D materials, doping emerges as a uniquely powerful control parameter, yet, conventional gating devices allow exploration of only a very limited range of doping that might preclude driving a topological transition in a metallic compound.

Transition-metal dichalcogenides (TMDs), particularly monolayer \NbSe{}, are ideal candidates to host topological superconductivity \cite{yuan2014,zhou2016,he2018}. Their broken in-plane inversion symmetry and heavy transition metal atoms generate a strong out-of-plane effective magnetic field. This so-called Ising spin-orbit coupling (SOC) not only protects superconductivity against in-plane magnetic fields far beyond the Pauli limit \cite{xi_ising_2016, engstrom_upper_2025}, but also naturally favors unconventional pairing by making each Fermi pocket effectively spinless \cite{hsu_topological_2017, Shaffer2020}. However, accessing a topological phase might require extreme shifts in the chemical potential of the single layer. Misfit layer compounds (MLCs) bypass standard limitations by sandwiching transition metal dichalcogenides (TMDs) between charge-donating rock-salt layers. In particular, in the MLC (LaSe)$_{1.14}$(NbSe$_2$)$_2$, the large work-function of the TMD compared to the rocksalt drives a huge charge transfer (up to $\sim0.6$ $e^-$/Nb) while  preserving the 2D character, band structure, and Ising SOC of the \NbSe{} layers \cite{leriche_2021, zullo_2023, samuely_2021, le_du_2026}.

We have recently shown that this extreme intrinsic doping completely suppresses the usual charge density wave (CDW) \cite{zullo_2024} while superconductivity persists. Tunneling spectroscopy suggests an unconventional order parameter instead of a simple $s$-wave pairing \cite{leriche_thesis_2019}. Furthermore, alloying La$^{3+}$ for Pb$^{2+}$ in the rock-salt layers of \misfit{} provides a highly precise chemical knob to continuously tune the doping. We demonstrated that this tuning successfully manipulates the CDW order from a $3 \times 3$ to a $2 \times 2$ modulation, down to a complete collapse \cite{le_du_2026_cdw}, paving the way to explore other collective orders, such as superconducting, as a function of doping.

Here, we exploit this unique tunability to track the superconducting state of \NbSe{} across a remarkably wide doping range. Using bulk SQUID magnetometry and local scanning tunneling spectroscopy (STM/STS), we reveal a strongly non-monotonic phase diagram marked by an abrupt collapse of superconductivity at a critical composition ($x=0.4$). This boundary sharply separates a conventional $s$-wave regime at low doping from a disorder-sensitive, unconventional state at high doping. Crucially, orientation-selective in-gap modes emerge at atomic step edges in the highly doped phase. These signatures demonstrate a doping-driven quantum phase transition to a seemingly crystalline-topological order. This establishes MLCs as a uniquely tunable platform to engineer and explore two-dimensional topological superconductivity.

\section*{Superconducting phase diagram}

\begin{figure}[t]
\centering
\includegraphics[scale=0.35]{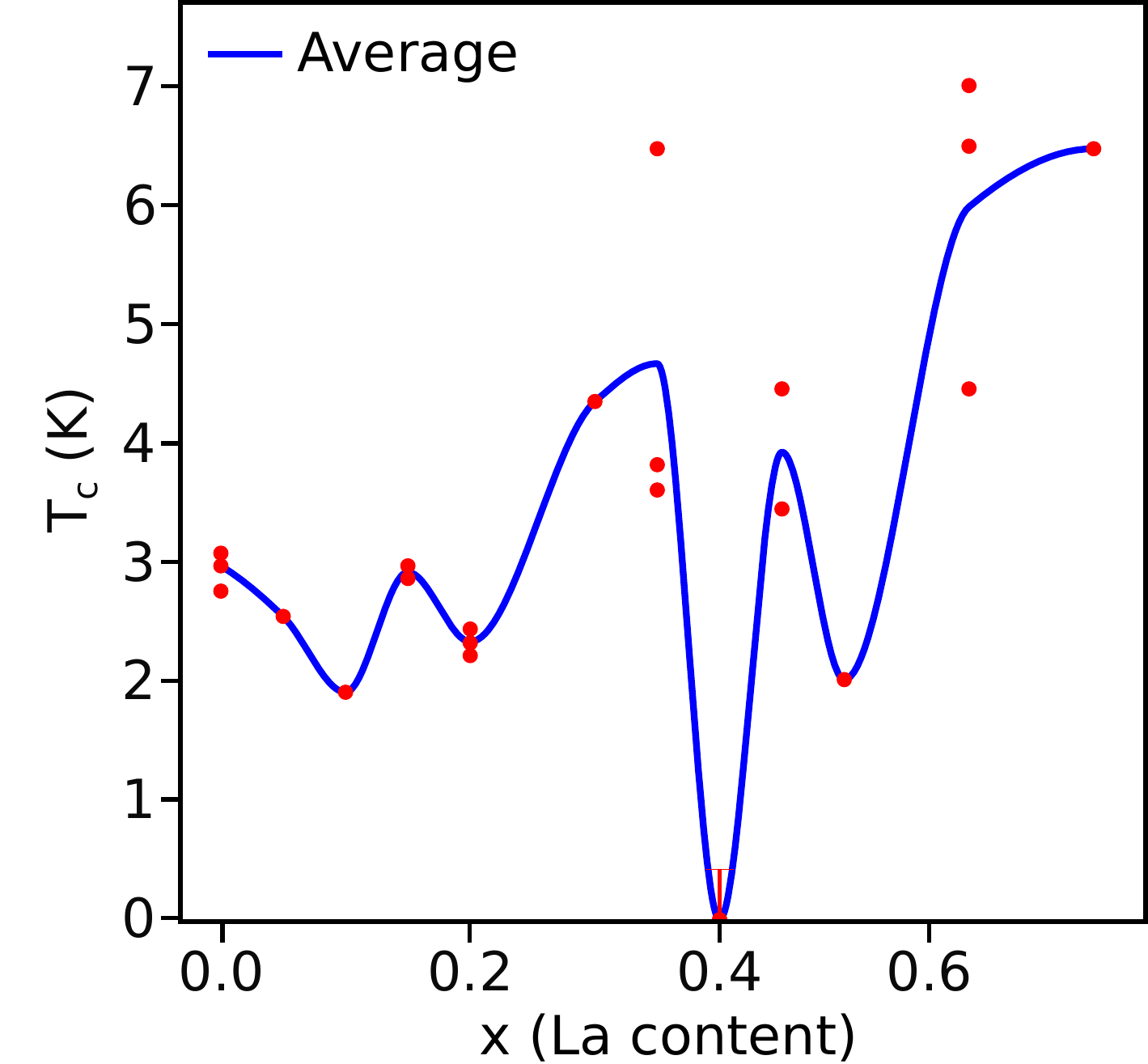}
\caption{\textbf{Doping-tuned superconducting phase diagram of \misfit{}.} $\Tc$ versus La content $x$: markers are individual samples, the curve is the $\Tc$ average. Superconductivity collapses abruptly at $x=0.4$, separating two superconducting regimes. Down-bars mark the base-temperature limit for non-superconducting samples. Representative magnetization curves and the $\Tc$ extraction are shown in Supplementary Fig.~S1.}
\label{fig:phase}
\end{figure}

To investigate the superconducting order parameter of the \misfit layer compound family, we first measured the superconducting transition temperature ($\Tc$) by SQUID-VSM magnetometry. We systematically extracted $\Tc$ as a function of the La/Pb ratio $x$ from the perfect diamagnetism under a small applied field (Supplementary Fig.~S1). The resulting phase diagram (Fig.~\ref{fig:phase}) is non-monotonic. In the low-doping regime ($0\leq x\lesssim0.35$), the system is superconducting with $\Tc$ fluctuating between $1.8$ and $4.4\,$K. At $x=0.4$, a sharp and reproducible suppression (observed across 6 samples) drives $\Tc$ below our $400$~mK base temperature, signaling a collapse of the superconducting state. Superconductivity re-emerges immediately between $x=0.45$ and $x=0.7$, forming a dome that peaks at the highest $\Tc$ of the series, near $x=0.7$.
 
Our experimental results naturally yield two questions: how can we explain this abrupt collapse at $x=0.4$, surrounded by two distinct superconducting regions? Could this suppression be merely driven by the evolution of the band structure, or by an interplay with the CDW order?

To answer this quantitatively, we solved the mean-field gap self-consistency equation as a function of chemical potential for all gap symmetries allowed by the $C_{3v}$ point group \cite{hanis_distinguishing_2024}, including a coupling to the CDW (Supplementary Sec. S2). Every symmetry yields a smooth, monotonic $\Tc(\mu)$, and the CDW coupling adds only weak modulations, none reproduces the sharp dip or the second dome (Figs.~S2-S3). A single self-consistent order parameter, even when including the CDW, cannot, therefore, account for the macroscopic observations. This phase diagram points to a change in the nature of the pairing across $x=0.4$, which bulk magnetometry, being intrinsically insensitive to pairing symmetry, cannot settle on its own. We therefore turn to a local probe.

\section*{Local collapse and change of gap shape}

To unveil the exact nature of these two superconducting regimes, it is imperative to probe the pairing symmetry directly at the atomic scale. Scanning tunneling microscopy and spectroscopy (STM/STS) provides a unique probe of the local density of states (LDOS), enabling us not only to track the evolution of the shape os the superconducting gap across the transition.

\begin{figure}[t]
\centering
\includegraphics[scale=0.35]{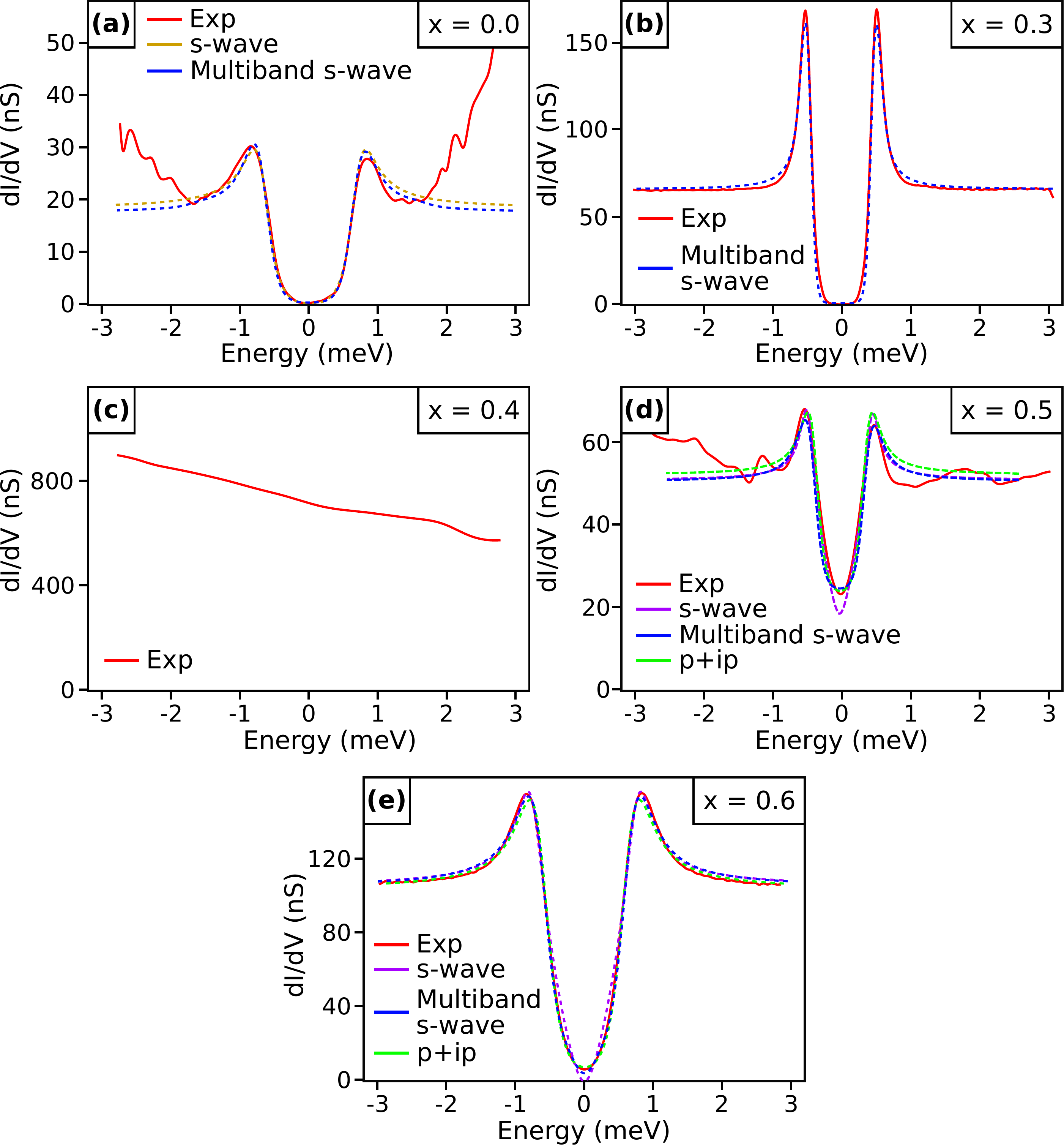}
\caption{\textbf{Local evolution and collapse of the superconducting gap.} Tunnelling spectra (STS, $350\,$mK, red) for $x=0.0,\,0.3,\,0.4,\,0.5,\,0.6$, overlaid with model fits (dashed) where a gap is present. The gap evolves from a fully-developed U shape at low doping, to a flat, featureless spectrum at $x=0.4$ where it is locally closed (no fit is shown for this composition), to a filled V shape at high doping. At low doping the spectra are reproduced by the multiband Schopohl-Scharnberg model. At high doping the nodal $f$-wave, chiral $p+ip$ and multiband models are mutually indistinguishable, so the line shape alone does not select a pairing symmetry. Extracted parameters are listed in Supplementary Table~S1.}
\label{fig:gap}
\end{figure}

Leveraging this local probe, we measured the local density of states by STS at $350\,$mK on five crystals spanning the transition ($x=0.0,\,0.3,\,0.4,\,0.5,\,0.6$). The local doping of each sample is independently verified by quasiparticle interference (the exact methodology is detailed in Supplementary Sec.~S3). The tunneling spectra measured on each sample are shown in Fig.~\ref{fig:gap} in red. A clear change of gap shape occurs across the transition. At low doping ($x=0.0$ and $x=0.3$), the spectra show a fully developed U-shaped gap, with the conductance dropping to zero around the Fermi level. At $x=0.4$, the gap is locally closed: the spectrum is flat and featureless (Fig.~\ref{fig:gap}c), confirming the collapse measured by SQUID magnetometry. Above the transition ($x=0.5$ and $x=0.6$), a gap reopens but with a qualitatively different shape. The gap appears V-shaped, with a sizable zero-bias conductance fluctuating from point to point across the surface.
 
To characterize the gaps at low doping $x<0.4$, we fitted the spectra with a two-band model (Schopohl-Scharnberg \cite{schopohl_scharnberg_1977}) appropriate to multiband \NbSe{}, in which a strongly paired $\Gamma$ pocket induces a gap in a weakly coupled pocket through interband scattering \cite{noat_2015,schopohl_scharnberg_1977} (Fig. \ref{fig:gap}a-b). This model reproduces the $x=0.3$ spectrum almost perfectly and the noisier $x=0.0$ spectrum well, supporting a conventional, multiband superconductivity in the low-doping regime, as expected for undoped, or weakly doped, \NbSe{}. The high-doping $x>0.4$ V-shaped spectra are qualitatively different, but their line shape alone is not discriminating: a nodal $f$-wave, a chiral $p+ip$, and even a multiband $s$-wave model incorporating a finite quasiparticle lifetime (as detailed in Supplementary Sec.~S4), reproduce the $x=0.5$ and $x=0.6$ spectra equally well (Fig.~\ref{fig:gap}c-d and Table~S1). Point spectroscopy thus cannot, by itself, unveil the symmetry of the high-doping state. To identify the exact gap symmetry, we now turn to spatially resolved spectroscopy. Specifically, we investigate the response of the gap to non-magnetic disorder and the emergence of in-gap states at atomic step edges.

\section*{Orientation-selective edge states}

\begin{figure*}[t]
\centering
\includegraphics[scale=0.35]{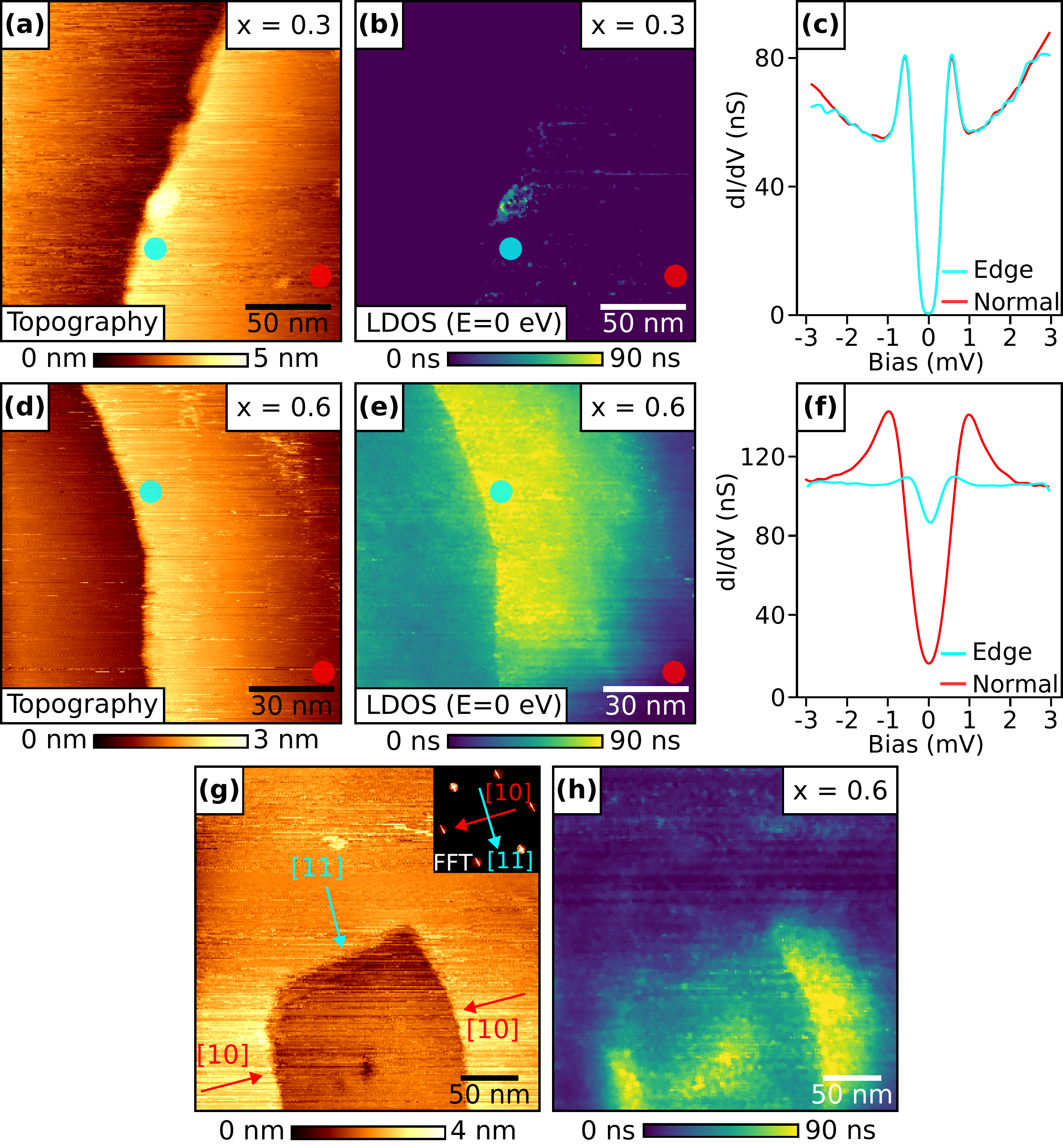}
\caption{\textbf{Disorder response and orientation-selective edge states.} (a-c) Low-doping sample ($x=0.3$): topography, zero-bias conductance map, and edge-versus-terrace spectra, the gap is uniform and disorder-robust. (d-f) High-doping sample ($x=0.6$): a sharp in-gap edge state appears at the step and non-magnetic defects locally suppress the gap. (g,h) Large-scale topography with lattice directions from the FFT (inset) and the corresponding zero-bias map: the edge state is present along $[10]$ and absent along $[11]$.}
\label{fig:edges}
\end{figure*}

To resolve the nature of these two distinct superconducting phases, we acquired spatially resolved spectroscopic (dI/dV) maps to probe the local density of states across natural atomic step edges in both doping regimes (Fig.~\ref{fig:edges}). This real-space mapping reveals a striking contrast. In the low-doping sample ($x=0.3$), the superconducting state is robust against non-magnetic disorder: individual tunneling spectra extracted at the step edge and on the flat terrace are indistinguishable, both exhibiting a fully depleted gap with absolutely no in-gap spectral weight (Fig.~\ref{fig:edges}a-c and Supplementary Fig. S5). Conversely, in the high-doping sample ($x=0.6$), the very same topographic features locally suppress the gap, and a massive accumulation of spectral weight emerges in the spectroscopic maps to form a sharp, well-defined state strictly localized at the step boundary (Fig.~\ref{fig:edges}d-f and Supplementary Figs. S6-S7). This abrupt switch from a disorder-robust to a disorder-sensitive behavior, combined with the emergence of bound states at non-magnetic defects, is a universal hallmark of a phase transition from a conventional to an unconventional order parameter \cite{frigeri_superconductivity_2004}.
 
To further probe the nature of this unconventional phase, we examined a natural step edge exhibiting a distinct rectangular shape, which locally exposes two perpendicular crystallographic boundaries (Fig.~\ref{fig:edges}g). Spectroscopic mapping along the boundary perpendicular to the dense atomic direction (the $[10]$ direction) reveals a prominent, sharp in-gap state. Conversely, when mapping the perpendicular boundary oriented at 90° (perpendicular to the $[11]$ direction), this localized spectral weight completely vanishes (Fig.~\ref{fig:edges}g,h). This striking observation demonstrates that the emergent boundary modes are orientation-selective. Such pronounced spatial selectivity provides a unique experimental fingerprint, allowing us to discriminate among the various allowed unconventional order parameters, as we will now discuss.

\section*{Bogoliubov-de Gennes analysis of the boundary modes}
To interpret the orientation selectivity, we computed the boundary spectra of candidate order parameters on finite \NbSe{} nano-ribbons. We employ a tight-binding description based on Density Functional Theory (DFT), considering three Nb $d$-orbitals ($d_{z^2}$, $d_{xy}$, and $d_{x^2-y^2}$) that are dominant at the Fermi level. We take into account both Ising and Rashba SOC. We tested all the possible pairing order parameter compatible with the representations of $C_{3v}$ symmetry \cite{hanis_distinguishing_2024} (Supplementary Secs.~S6-S7).

The nano-ribbon approach enables us to selectively expose different crystallographic edges and compute their respective boundary spectra.
A topologically protected edge state manifest as an in-gap band that crosses zero energy and link occupied and unoccupied bulk sectors. Trivial edge states can also give rise to in-gap states but they do not exhibit crossings near the Fermi level and thus do not connect both bulk sectors. 

There is a direct relationship between the symmetry of the order parameter and the edge states. For instance, time reversal breaking chiral order parameters, that are characterized by a Chern number $C$, manifest as dispersive protected edge modes, whatever the orientation of the edges. This is different for a time-reversal-crystalline symmetric superconductors whose topology is governed by a mirror-graded winding number $\nM$. This invariant relies on the three mirror symmetries of the $C_{3v}$ group. If the ribbon preserves globally the mirror symmetry of the crystal (Supplementary Fig. S8), then topologically protected edge states are present. Otherwise, if the ribbon globally breaks mirror symmetry trivial edge states with anti-crossing near Fermi energy are found (Fig. \ref{fig:edges}).

The allowed superconducting order parameters are classified by the three irreducible representations of the $C_{3v}$ point group of the crystal, which are distinguished by their spatial symmetries \cite{hanis_distinguishing_2024}. There are the two-dimensional $E$ representation (which enables chiral states), and the one-dimensional $A_1$ and $A_2$ representations which are respectively even or odd under the $C_{3v}$ mirror symmetries. The possible unconventional order parameters fall into three categories, namely chiral, nodal and crystalline symmetric. A chiral $E$-type state such as ($p_x+ip_y$) has a Chern number $C=1$ and produces dispersive modes that cross zero energy regardless of the ribbon termination (Supplementary Fig. S10). This orientation-independent signature indicates a time-reversal-breaking phase. This is not compatible with our experimental observations that display edge states only for some particular edges. The odd-parity $f$-wave states are nodal and host a flat, non-dispersive band at zero energy. Their nodal lines project onto different edges, so $A_2$ can produce zero-energy states along the observed $[10]$ direction. However, its flat band would manifest as a single, sharp peak fixed at zero bias. An analogy can be drawn with $d$-wave cuprate superconductors, where the relative orientation of the boundary and the nodal lines governs the formation of Andreev bound states. In cuprates, successive Andreev reflections at a [11] boundary create a dispersion-less flat band that yields a sharp zero-bias conductance peak, whereas a [10] boundary does not \cite{hu_1994, kashiwaya_2000, aprili_1999, krupke_1999}. By the same mechanism, a purely nodal $f$-wave state in our system would produce a sharp Andreev bound state peak, which contradicts the broad continuum we observe experimentally. These cases are shown in Supplementary Fig. S9. 

\begin{figure*}[t]
\centering
\includegraphics[scale=0.35]{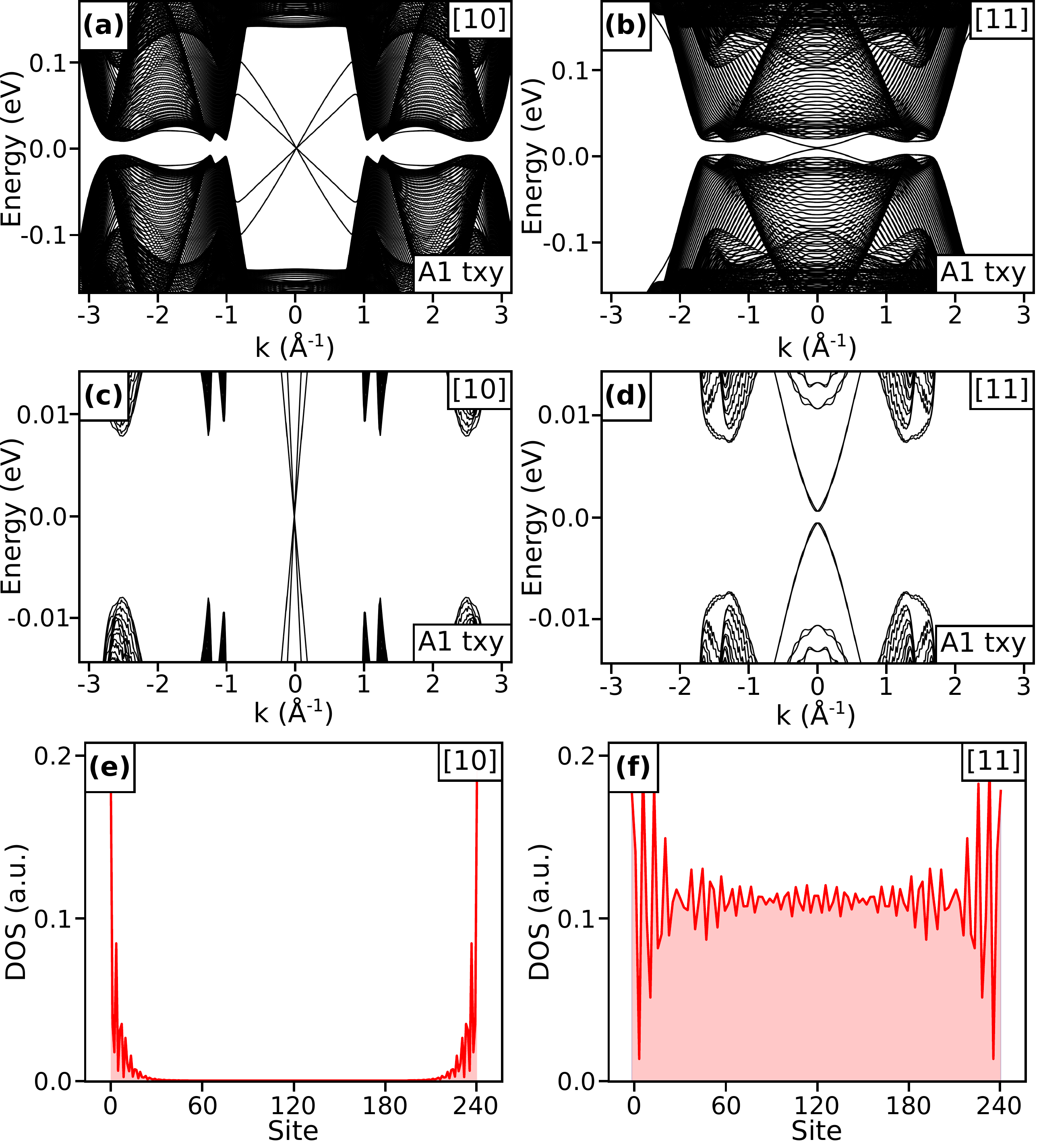}
\caption{\textbf{Boundary modes of the $A_1$ equal-spin triplet.} Real-space BdG nanoribbon spectra for the $A_1(t_{xy})$ state, which has bulk Chern number $C=0$ and mirror invariant $\nM=1$. (a,b) Band structure over the full energy window for the $[10]$ and $[11]$ terminations. (c,d) Zoom on the gap: on $[10]$ a pair of edge states disperses across the gap and crosses $E=0$ near $k=0$ (the helical, time-reversal-symmetric mode protected by the crystalline mirror symmetry) whereas on $[11]$ the spectrum remains gapped and the in-gap states approach but do not cross $E=0$. (e,f) Transverse weight of the in-gap states across the ribbon ($N=240$ sites, sites 1 and 240 are the edges): sharply peaked at both edges on $[10]$ and spread across the whole width on $[11]$. This is the signature of a genuine boundary mode versus delocalized finite-size states. The chiral $p+ip$ and nodal $f$-wave channels, which fail to reproduce this orientation selectivity, are shown in Supplementary Fig.~S8 and S9.}
\label{fig:bdg}
\end{figure*}

Let's now focus on crystalline symmetric order parameters for which we find that the equal-spin triplet belonging to the fully symmetric $A_1$ channel reproduces the data. This time-reversal symmetric triplet order parameter has an in-plane $\dvec$-vector with a Chern number $C=0$. It is invariant under the $C_{3v}$ mirror and characterized by a non-trivial mirror invariant $\nM=1$. This order parameter yields an helical dispersive mode that crosses zero energy along the mirror-preserving $[10]$ direction, while the $[11]$ direction remains gapped (Fig.~\ref{fig:bdg}b,d). The transverse profile of the in-gap states confirms this distinction. The $[10]$ mode is highly localized at the edge, whereas the $[11]$ spectral weight extends across the ribbon (Fig.~\ref{fig:bdg}e,f). This behavior is independent of the ribbon width over the range we computed ($80$-$240$ sites), confirming it originates from the order parameter rather than finite-size effects. Among the tested symmetries, the $A_1$ triplet is the only time-reversal-symmetric solution that reproduces the measured orientation dependence of edge states. A non-unitary $E$ triplet reproduces it as well (Supplementary Fig. S11), but breaks time-reversal symmetry.

\section*{Discussion}

Taken together, both SQUID magnetometry and scanning tunneling spectroscopy find a vanishing gap at $x=0.4$ in \misfit. Below $x=0.4$, conventional two-band superconductivity is found. It appears unaffected by non-magnetic disorder, as expected from Anderson's theorem for a conventional singlet $s$-wave order parameter. Above the $x=0.4$ doping level, the gap appears V-shaped and highly disorder-sensitive. It displays orientation-selective edge modes, which point towards a time-reversal-symmetric, crystalline-topological superconducting order parameter. We interpret the transition at $x=0.4$ as a doping-driven phase transition separating two distinct superconducting regimes. In the low-doping regime, a large Fermi surface and a high density of states screen the Coulomb repulsion, and electron-phonon coupling stabilizes conventional $s$-wave pairing \cite{xi_gate_2016,horhold_two-bands_2023}. Electron doping reduces the hole pockets and lowers the density of states. The weakened screening allows the residual repulsion to favor an unconventional pairing. \cite{wickramaratne_ising_2020,siegl_friedel_2025,gibelli_universal_2025}.

Identifying the exact symmetry above $x=0.4$ remains open. We present the $A_1$ triplet as a compatible candidate, without excluding other symmetries. Note that this order parameter is equal-spin in-plane triplet. It is not compatible with pure Ising SOC, it requires some additional Rashba SOC that is expected to be quite sizable in the misfit due to the large charge transfer between the TMD and the rocksalt layers.

Two alternatives to the $A_1$ triplet are worth noting. A finite-momentum (FFLO) state would be strongly disorder-sensitive and would lead to V-shaped gaps. A non-unitary $E$ triplet would reproduce the same orientation-selective edge mode without invoking Rashba coupling or finite-momentum pairing, at the cost of a gapless, time-reversal-breaking order parameter.

Two further considerations bound our interpretation. First, chiral symmetries are strictly defined only in two dimensions. In these three-dimensional misfits, interlayer coupling could hybridize a chiral state and suppress its edge mode along $[11]$, so a quasi-chiral $p$-wave scenario cannot be strictly ruled out. Second, the incommensurate rocksalt potential breaks the three-fold rotation of the \NbSe{} lattice and lowers the point group from $C_{3v}$ to $C_s$, so a strict $C_{3v}$ assignment should be treated with caution. This does not affect our conclusion, because the orientation-selective edge states probe the crystalline-topological character, which is protected by the single vertical mirror that $C_s$ retains rather than by the full $C_{3v}$ group.

In conclusion, our data establish that chemical doping in misfit layer compounds drives \NbSe{} through a phase transition with a superconducting order evolving from a conventional multiband regime into a topological superconducting phase. This establishes the \misfit{} family as a unique, bulk, and tunable material platform for exploring topological superconductivity. Unveiling the exact pairing symmetry of the high-doping state, will require further phase-sensitive experiments, such as quasiparticle interferences and Josephson effect.

\section*{Methods}

\small
\noindent\textbf{Crystal growth.} Single crystals of (La$_{x}$Pb$_{1-x}$Se)$_{1.14}$(NbSe$_2$)$_2$ were prepared by the solid-state reaction of the elemental precursors (i.e., La, Pb, Nb, Se) and subsequent chemical vapor transport using I$_2$. Under inert atmosphere, submillimeter-sized La powder was freshly scraped from the ingot (Strem Chemicals, 99.9\%) and mixed with Pb (Sigma Aldrich, 99.9\%), Se (Alfa Aesar, 99.999\%) and Nb powder (Puratronic, 99.99\%) in molar ratio La/Pb/Nb/Se = 1.14x/1.14(1-x)/2/5.14 for x = 0; 0.30; 0.40; 0.50; 0.60; 0.70. The mixture was manually ground in an agate mortar and transferred in a silica tube, which was subsequently evacuated to $10^{-3}$ torr and sealed by flame. For x = 0.30; 0.40; 0.50; 0.60; 0.70, the mixture was firstly heated to 200°C at a rate of 50°C h$^{-1}$ and held for 12h. Then the temperature was raised to 1000°C at the same rate and heated for 240h before the furnace was subject to radiative cooling. For x = 0, the sealed mixture was heated to 800°C at a rate of 50°C h-1 and held for 144h before the furnace was subject to radiative cooling. The lustrous black powder (ca., 1000 mg) obtained for each reaction was placed in the clean silica tube (length: 15 cm) together with 50 mg of iodine (Aldrich, 99.9\%), followed by the evacuation at liquid nitrogen temperature (77 K) and subsequent flame sealing. The reaction mixture was loaded into the furnace designed to create temperature gradient: the lump of the mixture (x = 0.3; 0.40; 0.50; 0.60; 0.70) was heated at 900°C on one side of the tube and the other side was held around 750°C. For x = 0, the lump of the mixture was heated at 800°C on one side of the tube and the other side was held around 700°C. After 240 h of the thermal treatment, the reaction mixture was removed of the furnace. The single crystals grown on the wall of tube were washed with water and ethanol and stored in vacuum. The compositional integrity of the obtained crystals was checked from their backscattered electron images and energy-dispersive X-ray (EDX) spectra acquired on scanning electron microscopy (JEOL JSM 5800LV).
(La$_x$Pb$_{1-x}$Se)$_{1.14}$(NbSe$_2$)$_2$ crystals were cleaved \textit{in situ} at room temperature under vacuum in a pressure of $P\approx 1 \times10^{-7}$mBar.
 
\noindent\textbf{SQUID magnetometry.} Bulk magnetization was measured in a SQUID-VSM in a small applied field (i.e. 50 Oe). $\Tc$ was defined from the intersection of linear extrapolations of the normal-state signal and of the diamagnetic drop (Supplementary Fig.~S1). Two modes were used, a slow $400\,$mK mode and a fast mode limited to $\sim2\,$K, for non-superconducting samples the error bar marks the lowest temperature reached.
 
\noindent\textbf{STM/STS.} STM/STS was performed with a home-built low temperature STM at a base temperature of $350\,$mK and under ultra high vacuum under a pressure of $P<1 \times10^{-10}$mBar. Differential conductance was acquired at a set-point current of $200\,$pA. Zero-bias conductance maps were taken inside the superconducting gap, step-edge orientations were extracted from the Fourier transform of atomic-resolution topographies.
 
\noindent\textbf{Gap fitting.} Spectra were fitted with a single-band angle-resolved BCS model with Dynes or self-consistent impurity broadening and the reciprocal-space two-band Schopohl-Scharnberg model, thermally broadened at an effective temperature. Full expressions, procedures and extracted parameters are in Supplementary Sec.~S4 and Table~S1.
 
\noindent\textbf{Mean-field and BdG calculations.} The self-consistent $\Tc(\mu)$ scan and the real-space BdG nanoribbon calculations (three $d$ orbitals, two spins, Ising and Rashba SOC, Chern and mirror-winding invariants) are described in Supplementary Secs.~S5.

\normalsize

\section*{Data availability}
The data that support the findings of this study are available from the corresponding author upon request.

\section*{Acknowledgements}

This work was supported by the French Agence Nationale de la Recherche through the contract ANR Misfit (ANR-21-CE30-0054), the ANR SURIKAT (ANR-23-CE30-0036) and the ANR MASCOTE (ANR-24-CE30-1342).

\section*{Competing interests}
The authors declare no competing interests.

\clearpage

\clearpage

\bibliographystyle{unsrtnat}
\bibliography{references}

\end{document}